# Real Time State Estimation of Power Grids Using Convolutional Neural Networks and State Forecasting Via Recurrent Neural Networks

By

Sahil Vohra

A Thesis

Presented To

The University of Guelph

# Abstract

Real Time State Estimation of Power Grids Using Convolutional Neural Networks And State Forecasting via Recurrent Neural Networks

Sahil Vohra

University of Guelph, 2021

Advisor

Dr. Hadis Karimipour

Power grids play a very important role in delivering electrical energy to homes, industries and other places that require it. Because of this increased demand they are facing a great challenge of voltage variations. This happens due to varied use of energy-consuming devices and appliances like electric vehicles, industrial consumption, occasional peak in energy demands etc. For these fluctuations in demands, it becomes extremely important to monitor the conditions at which the power grid operates. Once these conditions are known, the energy production can be manipulated to meet the demand. It has been found that the existing Power System State Estimation (PSSE) techniques may not be good in producing optimal Performance. Moreover, they are also expensive in terms of computational processing. To address this problem, this research proposes a state estimation method for power grids using Convolutional Neural Networks (CNN). It was found that the model produced an RMSE of $2.57 \times 10^{-4}$, which was comparatively accurate than one of previous studies involved in making the estimation using a Prox Linear Model ($2.97 \times 10^{-4}$).

Furthermore, the research also proposes Power System State Forecasting for improving system awareness and resilience. The forecasting is carried out using a model of Recurrent Neural Network (RNN). This model helps in accounting for long-term nonlinear aspects present in data and based on that it does the forecasting. The proposed model forecasted with a RMSE of $2.53 \times 10^{-3}$, which is comparatively equal to the previous study mentioned above ($2.59 \times 10^{-3}$).

# Acknowledgements

There are two people whom I would like to thank the most along this enlightened and fostering journey of this research work at the University Of Guelph.

The work done in this thesis could never be so appreciable without the help of my supervisor, Dr. Hadis Karimipour. I am always grateful to her for the support and constant feedback which helped me in finishing this research in the best possible manner. She has provided me with this great opportunity and has kept faith in me despite my mere understanding of the subject. I would also like to thank Shahrzad Hadayeghparast, PHD student of Dr. Karimipour, for always helping and supporting me throughout the work. Her teachings and supportive materials have immensely helped me in developing myself and achieving better results for the research work.

My sincere thanks is also due to my family and friends for believing in me and my hard work. Especially my parents, who always encouraged me with love and support to progress in my life and do better without much restrictions or worries of any kind.

# Content

# List of Tables

# List of Figures

# Chapter 1: Introduction

One of the main functions of a power grid is to connect a variety of electric power generators to different customers via a line of transmission along with distribution of a network of power grids throughout a large geographical region. In such a situation the reliability of the power grids becomes a crucial factor. For instance on 14th of August, 2003 a significant area in the northern and midwest US along with Ontario in Canada experienced a larger electric breakout. This impacted the area whose population was about 50 million people. The total loss in this was estimated to be around $14 billion in the United States and around $2.3 billion in Canada.

System monitoring is one of the essential things because it helps us to ensure the reliable and continuous operations of power grids. It shows related information in the states and variables affecting the power grids based on the recordings of the meters that could have been placed on some important elements and components of the power grid, for example substations. As for the meters, they may include voltages of all the buses, reactive and real time power injections, and also the reactive powers of various branches that flows throughout the subsystem of a particular power grid. Typically these measurements are transferred into a control center, where the staff of the control center along with the assistance of computers collect some important data and generate a centralized control and monitoring system for enhancing the capabilities of a power grid. These measurements are usually stored in a system call as Supervisory Control And Data Acquisition (SCADA) system

The purpose of system estimation in the use of system monitoring is to estimate the states of the power grid via which the analysis of measurements and power system is performed. The process of state estimation is to estimate the unknown state variables of a power grid based on measurements from the meters. The staff from the control center can use the state estimations as they conduct contingency analysis, where they can reason out the potential problem related to the operation of the power grids, certain actions they can make to avoid these issues, and the related side effects of these actions. As an example, they can opt to increase or decrease the yield of generators for maintaining smooth functioning and operation even when faults like generator breakdown are present.

North American power grid system is one of the most significant achievements of this century as it's a cyber-physical system which has capabilities of transmission and a distribution system of infrastructure which could deliver electricity from power generators to the consumers [**1**]. The modern day power grids are facing a big challenge in terms of sudden and unprecedented variations of levels of loads and fluctuations of voltages. This is inline with the growing number of the Electric vehicles and the distributed renewable generators. with the growing deployment of EVs and distributed renewable generators. As a result the growing market share kept on rising for EVs, this received a record of 1:13% just in the year 2017. In a nutshell, a total of around 1 million EVs were traded in the US alone from 2008 to 2018 [**2**]. With such an annual growth of 10–60% from the year 2004, the consumption of renewable energy already reserved 19.3% of the overall consumption of energy in the world in 2017.

In order to code with these challenges that the huge integration of EVs and renewable generators have brought about, the Department of Energy in the United States aims to improve existing grid systems and modernize them by development of advanced electronics, communication along with measuring instruments [**3**]. The general opinion is that for truly transformation of power grid networks into something reliable and substantial infrastructure [**4**]. As a result, there is a development of scalability and algorithms that are robust enough for grid states and topology infrastructure interference, as well as for associating various resources of crucial importance [**4**, **3**].

## 1.1. Motivations

Since the equations governing the nonlinear nature of the flow of power within the electrical power grids are concerned, there have been proposed numerous approaches for state estimation and for forecasting. Also due to the increasing amount of nonconvexity, the existing power system state estimation (PSSE) system becomes much more difficult in terms of computational expenses and can sometimes yield non optimal performances.

That is why there is a need for system state estimation which can help us to determine states even when there are problems mentioned above. Amd also a statem state forecasting method that can help to determine states even ahead of the time horizon.

### 1.1.1. State Estimation

There is an access use of supervisory control and data acquisition (SCADA) elements in most of the electrical appliances throughout the power grid network for supporting all the computer systems. This collects the relevant data and can be used for numerous relevant applications monitoring a system, operating an economic system, assessment of security tasks, power generation. For making any assessment or an action of control it is important to make an estimate which is reliable. For this reason, the amount of physical measurements are not supposed to be limited to the quantities that are required for the calculations of the flow of power. Moreover, there are also errors with more than one physical quantity that can lead to unnecessary and suboptimal results.

## 1.1.2. State Forecasting

For many researchers, this purpose of performing state forecasting is mostly carried out by techniques like Kalman filter, approaches related to moving horizon and first-order vector autoregressive (VAR). [**5**, **6**] One of the problems with all the above state predictors is that they assume linear dynamics. However in real life practice, the current state is dependent on the previous or estimated state. This dependence is non-linear and cannot be accurately characterized.

Non-linear State forecasting methods were proposed in where a Feed Forward Neural Network (FNNs) based State prediction has been proposed with transition mapping which was modeled by a single hidden layer. Unfortunately the total number of parameters in Feed Forward Neural Network tends to increase with the increase in input sequences. This would make FNNs inappropriate for considering long-term dependencies in the series of voltage and time

Deep recurrent neural networks (Deep RNNs) are capable of working with fixed number of parameters even if they are made to work with variable length of input sequences. Another thing that makes them suitable for this work is that they are able to capture very complex and non-linear dependencies that may be present in time series.

## 1.2. Objectives

In this thesis, the main research objective is:

To study the state estimation of the power grid variables and determine a deep learning convolution algorithm that could be utilized in state estimation with great accuracy. In addition to this, the research also aims to develop a state forecasting method using deep recurrent neural networks for predicting even ahead of the time horizon.

This thesis aims to deliver an efficient method for state estimation in smart power grids. This can be achieved by conducting a literature survey and some experimental analysis. For accomplishing this goal of the thesis, following contributions were listed.

## 1.3. Organization

Following is the organization of the thesis:

*Chapter 2* provides the background information for the thesis project. It introduces state estimation and forecasting definitions, different types of errors, explanation about the convolutional neural network and deep recurrent neural network, learning process of RNN etc.

*Chapter 3* mentions the methodology of the research work. State estimation was carried out by altering the number of neurons and then the number of epochs. While state forecasting was carried out differently by trying out different structures of RNN networks.

*Chapter 4* describes the results and makes a general discussion on how those results were obtained. It describes graphs of the state estimates and forecasting results. It mentions the process of selection of the best and optimal approaches for both state estimation and forecasting.

Chapter 5 finally gives a conclusion for the project and mentions the possible future work for it.

# Chapter 2: Literature survey

State estimation is one of the most important aspects of any power system analysis. For operating any power system one can control it from a centre which is called load dispatching centre, where one can deduce different information about the present state of that particular power system. The categories of this information includes a variety of readings from meters, tap positions of transformers, states for any circuit breaker and the topology of the network. With regards to the transmission of all their information into a SCADA system, it is not always reliable. For instance, all the errors that are caused due to inadequate connections within transducers, data loss when transmission or any sort of malfunctioning measuring tools. So when any of these incorrect is used for any contingency analysis of a power system, we may find that the system gets wrong signals. This can also lead to irreverent tipping of the signals. For this reason, for determining the efficiency of the measurements even with the bad data or sometimes even loss of data, we find that the use of power system state estimation makes a big difference by making the system better. For classical approaches, the state estimation is typically performed with the measurements obtained from the SCADA system. In addition to this, Phasor Measurement Units (PMUs) can also potentially measure system's states and also considerably uplift the accuracy of the estimations of the states. However, PMUs are usually associated with higher prices and it might be little impractical to implement

PMUs at all the buses. For this reason there is a need to develop an algorithm that could be based on SCADA measurements.

## 2.1. Types of State Estimation

In terms of the variations in states with respect to time, other invariant behaviors of measurements, and also the static or dynamic model of the system which could be utilized during the operation, the state estimation could be categorised into three types:

1. **Static state estimation:** It can be defined as an algorithm for processing data for transforming all the redundant measurements and other information into a vector which is the state estimation, while the data which is measured is considered to be variable with time and the model of the state for the system is considered [**7,8,9**].

2. **Tracking state estimation:** It is also an algorithm which is based on one of the extensions of the state estimation method discussed above. This method utilizes any recent values that could be available from the system states which is then updated to an estimated value.This estimation happens non iteratively while the corresponding sampling period is running. This category of estimators have been derived from the use of natural need for generating static state estimators in an efficient manner as much as possible. This is in reference to the computational speed for making them better suitable for the implementation in real time. [**10**]

3. **Dynamic state estimator:** The way in which this estimator is different from the previous ones is that in addition to the current states, this estimator also utilizes the previous state estimates. This ability of forecasting the vecorts even ahead of the time horizon one step ahead is one of the most important advantages of the dynamic estimators. This kind of prediction can also give a longer time for making decisions for the system operators. This is due to the fact that it would allow things like dispatching economic related things, assessment of risks, and other related tasks can be performed before the execution. In this estimation, a model which is dynamic in nature is used for the timely nature of system states. On the other hand simple tracking or static state estimation may not require us for any modelling of a dynamic system.

# 2.2. Classification of Errors

Various analysis of observability shows us the number of measurements that are used in estimating the state vector. All state estimators determine the magnitude of voltage along with the phase angle of every bus from the measurements that are available from the real time recordings. This generally consists of the reactive and real power. Considering the severity of different errors that they introduce into the measurements, they can be categorized as follows [**11**].

1. **Extreme error**: In this error the absolute difference from the measured value to the real or true value goes beyond the value of $20\sigma$; $\sigma$ here is the standard deviation for measurements.

2. **Gross error (Bad data)**: In this error the absolute difference in measured and true value is in the range of $5\sigma$ and $20\sigma$.

3. **Normal error**: In this error the absolute difference from measures to the true value is less than $5\sigma$.

All the outliers and extreme errors in data are usually rejected by any prefiltering methods from the measurements. Some very simple checks like limit checks on the measurement data which are incoming can be used for eliminating various kinds of bad datas. This process there is an exception for the gross errors that may pass through this prefilter and may also get utilized in the process of state estimation. Hence their effect on the state prediction must be optimized. These gross errors can be generated by any kind of malfunctioning of the measuring instruments or could be due to a breakdown in the functioning equipment. This processor of bad data utilizes this process in detecting the effect of bad data throughout the measurements. All there are the estimations from the data obtained from measurements that were to be utilized for the state estimation.

# 2.3. Least-Absolute-Value Estimation

Let us assume a network of a power system that would be composed of N number of buses which could be made into a model graph of G := {N, L}, here N := {1, 2, . . .,N} consisting of the whole bus list, and L := {(n, n)} ∈ N ×N can be connected to all the lines; see Figure 1.

For every bus n ∈ N, let $V_n := v^r_n + jv^i_n$ to be denoting the respective complex voltage along with its magnitude $|V_n|$, and $P_n$ ($Q_n$) to be denoting injection of the active/reactive power. For each line (n, n') ∈ L, assume $P^f_{nn'}$ ($Q^f_{nn'}$ ) to be denoting the active /reactive power flow which could be seen at the point of 'forwarding' end. Also, assume $P^t_{nn'}$ ($Q^t_{nn}$) to be denoting flow of active/reactive power at the point of terminal end.

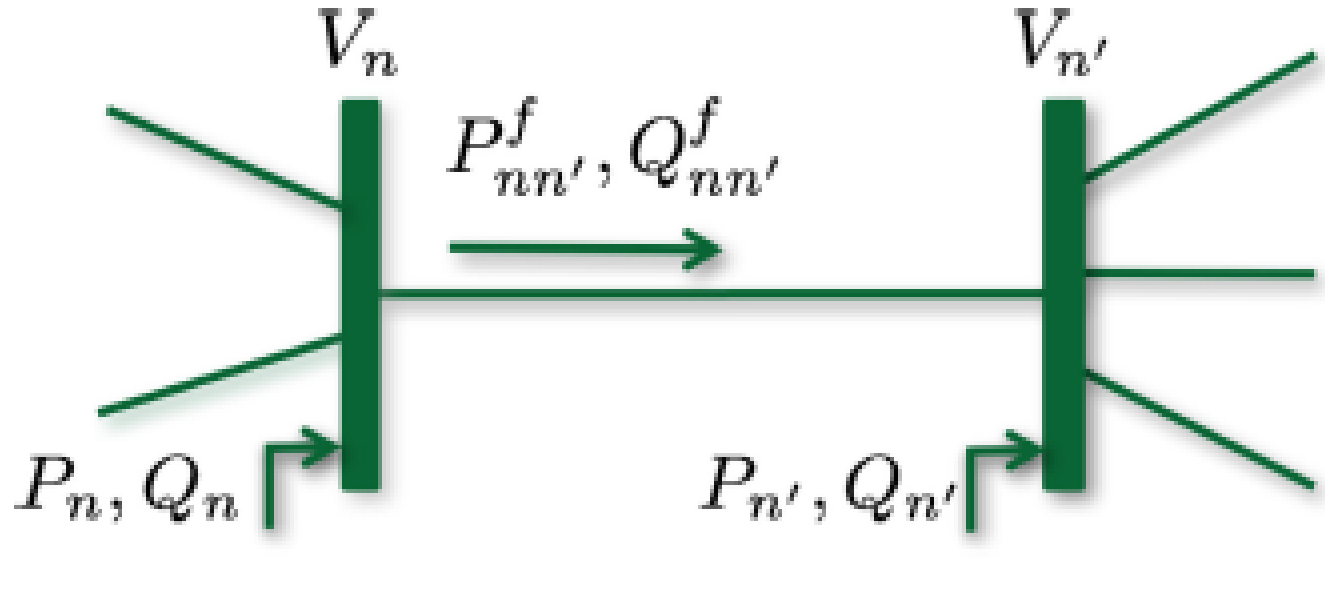

Figure 1: Connection between n and n' via line (n, n').

For performing any power system state estimation, assume $M_t$ to be total variables for the system that are measured at a given instant of time t.

Representing in a compact format, let $z_t := [\{|V_{n,t}|^2\}_{n\in Not}$ , $\{P_{n,t}\}_{n\in Not}$ , $\{Q_{n,t}\}_{n\in Not}$ , $\{P^f_{nn,t}\}_{(n,n)\in Eo}$ , $\{Q^f_{nn,t}\}_{(n,n)\in Eot}$, $\{P^t_{nn,t}\}_{(n,n)\in Eot}$ , $\{Q^t_{nn,t}\}_{(n,n)\in Eot}]^T$ to be the vector of measurement which would be collecting all the available quantities that would be measured at the time interval t. Here sets $N^o{}_t$ and $E^o{}_t$ are indicating the locations respective to the nodal and line quantities.

For any time instant t, the power system state estimation aims to recover the vector os system state: $v_t := [v^r{}_{1,t}\ v^i{}_{1,t} \cdot\cdot\cdot\ v^r{}_{N,t}\ v^i{}_{N,t}] \in R^{2N}$ into rectangular coordinates from a noisy $z_t$ measured values.

For any given measurement $z := \{z_m\}^M{}_{m=1}$ and a matrix $\{H_m \in R^{2N\times 2N}\}^M{}_{m=1}$ considering the below physical model.

$$z_m = v^T H_m v + \epsilon_m, \qquad \forall_m = 1, \ldots, M$$

Our objective is to recover $v \in R^{2N}$, where $\{\epsilon_m\}^M{}_{m=1}$ will account for all the noise measurement and model inaccuracies.

Considering the LAV criteria which is known to be very robust to the outlier:

$v^o := \arg_{v \in R2N} \min 1/M \quad \Sigma^M_{m=1} \mid z_m - v^T H_m v \mid$

Considering this, various solvers have been developed [**21**], [**22**]. Specifically a recent prox-linear solver was developed in [**23**] and it has some great advantages, this includes faster local quadratic convergence and efficient in dealing with unsmoothness and nonconvex nature in [**21**].

# 2.4. State Estimation Using Convolutional Neural Network

In addition to input and output layers, any ordinary Convolutional Neural Network (CNN) would have other layers like convolutional layers, batch normalization layers, activation layers, dropout and more.

## 2.4.1. Convolutional Layer

A typical convolution layer would consist of many filters, each one of them would be processing some data along with a convolution operation but only for its respective field.

Among many charactics of CNN, some of them are: sharing of parameters and sparse interactions [**12**], reducing quickly the free parameters that were learned and thus reducing the problem of overfitting and reducing dimensionality, that is carried out by some networks that are fully connected. In addition to this, byt adding GPU acceleration we can have a much faster computation of CNN, which could be highly applicable for TSA scenarios.

## 2.4.2. Activation Layer

This layer helps in introducing the neural networks with any form of nonlinearity. A typical activation layer would introduce nonlinearity and would be much helpful in controlling the flow of information from the input and output terminals. One of the activation functions, rectified linear unit (ReLU) [**13**] is generally considered over some other activation functions. This is due to the fact that it is one of the non saturating functions and can produce an acceleration in training of the neural net 6 times faster without posing any great threat to the generalization ability of the neural net [**14**].

## 2.4.3. Batch Normalization

A typical Batch normalization [**15**] would mainly be owing to the steps that would be used for fixing the means and for each layer their input variance. With regards to the effects of doing batch normalization, it can help significantly in improving the speed for training. It also helps in improving stability for a neural network. Many experiments have demonstrated that one of the reasons for this is the effectiveness that the normalization process creates [**16**]. Overall, the process of batch normalization always lightens the difficulties for training any typical deep neural networks. Also it promotes the deep learning method to a developed state.

## 2.4.4. Dropout

One of the simple and extremely effective methods for minimizing the problem of overfitting is dropout. It randomly drops certain units from the deep neural network during the training phase [**17**]. This helps in preventing high co adaption.

It is due to the existence of these new features like ReLU activation function or Batch Normalization, it became possible to train the deep neural networks directly without any need for a tedious work of any layerwise unsupervised pre training as explained in some of the previous literatures [**18, 19, 20**]. Hence, this makes the model training much more suitable and convenient for certain applications of TSA, which might require such faster training and updating methods.

# 2.5. State Forecasting With Deep Recurrent Neural Network

So far the power system state forecasting has been made by (extended) Kalman filtering, first-order vector autoregressive (VAR) modeling and moving horizon approaches. [**5, 6**] One of the problems with all the above state predictors is that they assume linear dynamics. However in real life practice, the current state is dependent on the previous or estimated state. This dependence is non-linear and cannot be accurately characterized.

Non-linear State forecasting methods were proposed in where a Feed Forward Neural Network (FNNs) based State prediction has been proposed with transition mapping which was modeled by a single hidden layer. Unfortunately the total number of parameters in Feed Forward Neural Network tends to increase with the increase in input sequences. This would make FNNs inappropriate for considering long-term dependencies in the series of voltage and time

Deep recurrent neural networks (Deep RNNs) are capable of working with fixed number of parameters even if they are made to work with variable length of input sequences. Another thing that makes them suitable for this work is that they are able to capture very complex and non-linear dependencies that may be present in time series.

## 2.5.1. What Is A Deep Recurrent Neural Network

Any system state estimation system would require forecasting for dealing with missing entries and obtaining system awareness even ahead of time which is required as such problems can arise many times in a SCADA system do do some communication or transfer failures. In this section a Power System State Forecasting method is proposed to predict the upcoming state $v_{t+1}$ at time t + 1 from the available time series states $\{v_\tau\}^{t}_{\tau=0}$. The analytical steps of estimation and prediction are as follows:

$$v_{t+1} = \varphi(v_t, v_{t-1}, v_{t-2}, \ldots, v_{t-r+1}) + \xi_t \qquad (1)$$

$$z_{t+1} = h_{t+1}(v_{t+1}) + \varepsilon_{t+1} \qquad (2)$$

where $\{\xi_t, \varepsilon_{t+1}\}$ are accounting the inaccuracies in the model; and φ Is the unknown and nonlinear function that is used to capture the transition of the states. while $h_{t+1}(\cdot)$ is The function for accounting measurements $z_{t+1}$ at time slot t+1. In order to perform the state forecasting, what is required for the function φ, to estimate or approximate, is a task that will make use of RNN modelling. The parameter $r \geq 1$, is used for the total number of lagged states that are used to predict $v_{t+1}$.

## 2.5.1. Learning Process

RNNs are Neural network models that are designed to learn from a series of data which has a correlation of time in between them. Unlike FNNs, RNNs Have the capability of killing long-term memory inputs that could entail large sequences of r. They are also capable of processing variable input sequence length..

For the input sequence $\{v_\tau\}^{t}_{\tau=t-r+1}$, and initial state $s_{t-r}$, an RNN determines the hidden state vector $\{s_\tau\}^{t}_{\tau=t-r+1}$ through repetition

$$s_\tau = f(R^0 v_\tau + R^{ss} s_{\tau-1} + r^0) \qquad (3)$$

where f is the activation function accounting nonlinearity (e.g., a ReLU or sigmoid unit), and the vector $r^0$ and the coefficient matrices $R^0$, $R^{ss}$ are accounting the time-invariant weights.

## 2.5.3. RNN Layers

Deep RNNs are an extended version of RNNs having multiple layers for processing. Deep RNNs have the ability to learn very compact and complex representations of the series of time using a hierarchical and nonlinear transformation. It is because of this reason that the deep RNNs have numerous applications of processing sequences. This includes music prediction and crypto currency prediction. These areas have seen significant improvements after the use of deep RNN models. By making the use of multiple recurrent hidden layers and topping them on one another, one can produce the Deep RNN as follows:

$$s^{l}_{\tau} = f(R^{l-1}s_{\tau}^{l-1} + R^{ss,l}s^{l}_{\tau-1} + r^{l-1}, \;\; l \geq 1 \qquad (4)$$

Here l is the index for layers, $s^{l}_{\tau}$ is the hidden state of the l-th layer at time slot τ.

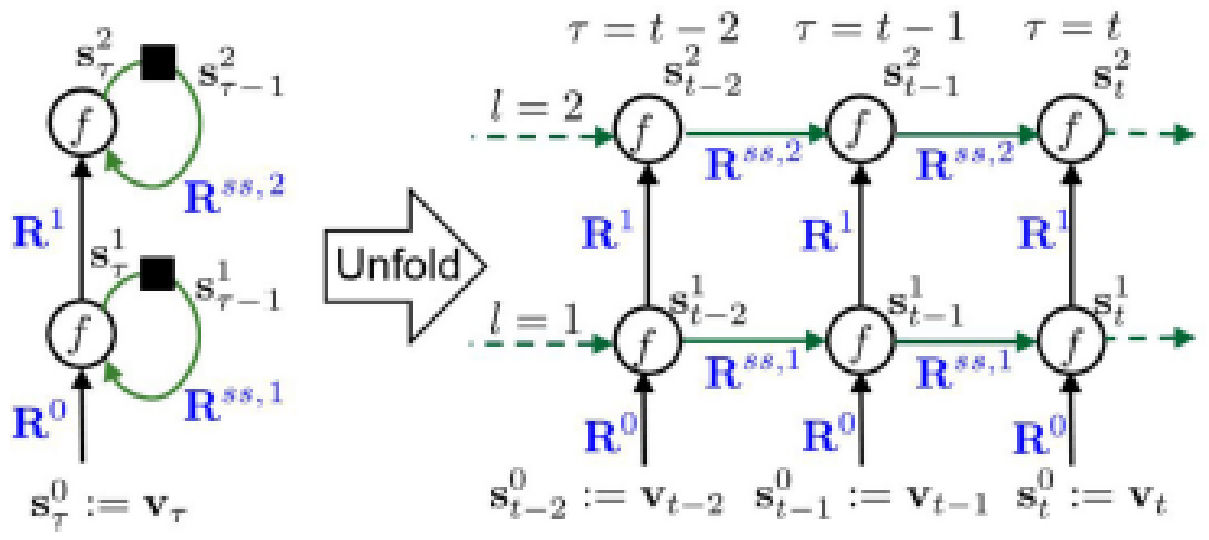

Figure 2: Deep RNN, Source [32]

Fig. 1 (left) represents the order of computing for equation (4) and for l = 2, considering the the bias vectors rl = 0, ∀l. After unfolding the graph and breaking the loops we get the deep RNN in Fig. 1 (right), in which the rows are representing layers, and columns are for time slots.

## 2.5.4. RNN Output

With regards to the output of the RNN, it can come in a variety of forms, there could be only one output per time slot or there could be a number of outputs or there could be a single output after a significant number of steps. The output of the RNN proposed in this paper is given by:

$$v'_{t+1} = R^{out} s^{l}_{t} + r^{out}$$

where $v'_{t+1}$ is the forecasted state of $v^{t+1}$ at time t, and ($R^{out}$, $r^{out}$) have all the weights for the output layer. Considering a history of a series of voltage and time the weights {$R^{out}$, $r^{out}$} and {$R^{l}$, $R^{ss,l}$, $r^{l}$} can be determined using the method of backpropagation.

# Chapter 3: Methodology and Considerations

## 3.1. State Estimation Using Convolutional Neural Network.

The data for training and testing was collected from the 2012 Global Energy Forecasting Competition (GEFC), this data can be found at https://www.kaggle.com. The evaluation of the estimation was made using normalized RMSE method: $||v' - v||^2 /N$, where v' is the ground truth as mentioned in the data and v is the predicted result from the neural network model. The model was made to perform using Google colab with a GPU hardware accelerator. A significant amount of work was also done on the data preprocessing part. First, the imported data was extracted into 'inputs' and 'labels'. Then, the whole data was also split into training and validation sets, in which 80% of data was used into training and the rest in validation. Since the data was scaled, there was also a requirement for descaling both voltage and phase angles. The data was then also normalized separately for both voltages and phase angles. Lastly, because of the requirements of Convolution Neural Network (CNN), the data also needed reshaping. Since the data was normalized during the preprocessing part, there was also a need for denormalization of the predicted results to be compared with labels of the validation set using RMSE.

Following are the data parameters:

| | |
|---|---|
| Number of features: | 490 |
| Number of classes: | 236 |
| Number of examples in the training set: | 14822 |
| Number of examples in the test set | 3706 |

Table: 1 List of data parameters

A CNN model was created with two hidden layers and an output layer. Both the hidden layers had equal numbers of neurons. The total number of neurons in the hidden layers (N) was varied and a total of 7 different models were created. The one that obtained the least value of RMSE over the validation/test data set was considered as the best model for prediction. All the models used ‘Adam’ as their optimizer and used ‘mae’ to measure error during convergence. Since each data set was linear or one dimensional, the model used ‘Conv1D’ layers from tensorflow for the hidden layers. The activation function was ‘Relu’ for both the layers. As for the output layer, a ‘Dense’ layer with no activation function was used to generate an output of dimension of (1, 236). Another machine learning technique, K-Nearest Neighbourhood (KNN), was also used to make predictions. Here, the value of 'k' was made to vary from 0 to 9 and the RMSE was observed for all of them

## 3.2. State Forecasting Using Deep Recurrent Neural Network

The data for training and testing was collected from the 2012 Global Energy Forecasting Competition (GEFC), this data can be found at https://www.kaggle.com/c/global-energy-forecasting-competition-2012-load-forecasting/data. The evaluation of the forecasting was made using normalized RMSE method: $||v' - v||^2/N$, where v' is the ground truth as mentioned in the data and v is the forecasted/predicted result from the model. The model was made to perform using Google colab with a GPU hardware accelerator. A separate file consisting of all 7 different models was created and imported to the Google colab file.

A significant amount of work was also done on the data preprocessing part. First, the imported data was extracted into 'inputs' and labels, From both of them only the labels were used for forecasting as they have values only for voltages and phase angles. Then, the data was split into training and validation sets, in which 80% of data was used into training and the rest in validation. Since the data was scaled, there was also a requirement for descaling both voltage and phase angles.

The data was then arranged in a way that 10 of the voltage and phase angle sequences were feeded into the model as the inputs and a single sequence right after those 10 was predicted. In other words, if sequences from i to i +10 were used as inputs then the predicted one will be equivalent to the i + 11 sequence. All the models were made to run for 200 epochs using the same training data. Then, RMSE was calculated and two plots were created. The first plot shows the variations in the voltage magnitude for a single time slot of t = 200 considering all the 118 buses. The second plot shows variation of all the values of phase angle of all buses at the same time. All the models used ‘Adam’ as their optimizer and used ‘mae’ to measure error during convergence.

# Chapter 4: Results and Discussion

## 4.1. State Estimation Using Convolutional Neural Network

### 4.1.1. Selection For The Number Of Neurons In Layers

A total of 7 different models were created with a varying number of neurons of hidden layers from 185 to 230 with increment of 5 for every new model. A separate model with 236 neurons was also tested. This was because the length of each prediction or label is also 236 i.e. 118 voltages and 118 phase angles.  Each model was made to run for 400 epochs. The RMSE for all these models are mentioned in the table below:

| **Hidden Layer Neurons** | **RMSE** |
|---|---|
| 185 | $3.37 \times 10^{-4}$ |
| 190 | $3.26 \times 10^{-4}$ |
| 195 | $3.27 \times 10^{-4}$ |
| 200 | $3.25 \times 10^{-4}$ |
| 205 | $3.32 \times 10^{-4}$ |
| 210 | $3.24 \times 10^{-4}$ |
| **215** | **$3.13 \times 10^{-4}$** |
| 220 | $3.30 \times 10^{-4}$ |
| 225 | $3.55 \times 10^{-4}$ |
| **230** | **$3.14 \times 10^{-4}$** |
| 236 | $3.42 \times 10^{-4}$ |

Table 2: List of Numbers of neurons in hidden layer with their respective RMSE values

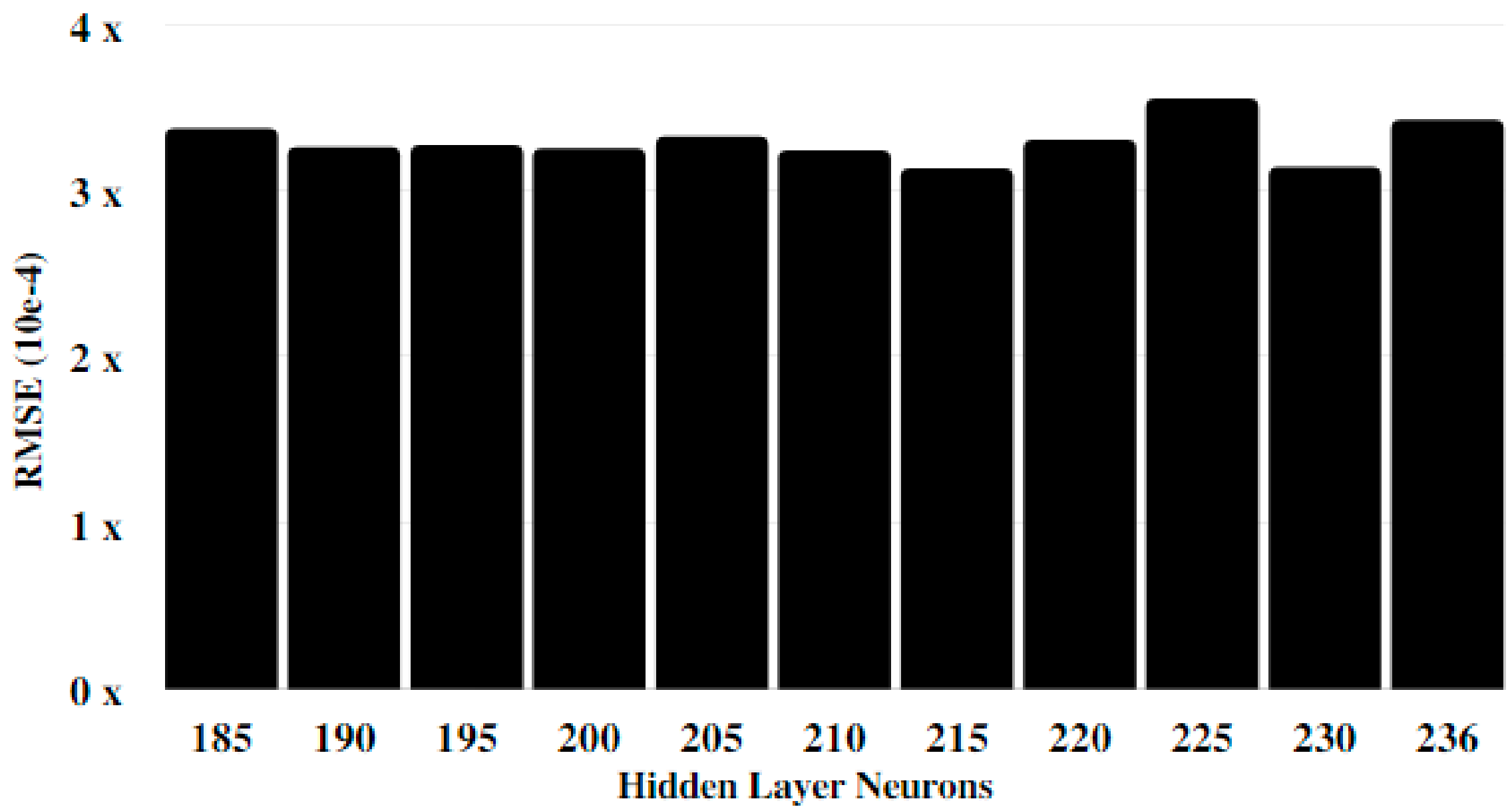

Figure 3: Numbers of neurons in hidden layer with their respective RMSE values

Out of all these models the model with 215 neurons in the hidden layer was having the least RMSE value of 3.13 x $10^{-4}$. However, the model with 230 neurons also had a very close RMSE (3.14 x $10^{-4}$). Hence both the models were further taken for consideration. Now with this model the total number of epochs were changed from 200 to 700 with an increment of 50 with every new iteration. Hence, the model was tested for 9 times with varying numbers of epochs. The results for each execution is mentioned in the table below.

## 4.1.2. Selection Of The Total Number Of Training Epochs

| Epoches | RMSE | |
|---|---|---|
| | **215** | **230** |
| 200 | $4.21 \times 10^{-4}$ | $4.57 \times 10^{-4}$ |
| 250 | $4.07 \times 10^{-4}$ | $4.00 \times 10^{-4}$ |
| 300 | $3.97 \times 10^{-4}$ | $3.93 \times 10^{-4}$ |
| 350 | $3.65 \times 10^{-4}$ | $3.73 \times 10^{-4}$ |
| 400 | $3.13 \times 10^{-4}$ | $3.14 \times 10^{-4}$ |
| 450 | $3.24 \times 10^{-4}$ | $3.15 \times 10^{-4}$ |
| 500 | $3.03 \times 10^{-4}$ | $3.08 \times 10^{-4}$ |
| 550 | $3.02 \times 10^{-4}$ | $3.26 \times 10^{-4}$ |
| **600** | $2.57 \times 10^{-4}$ | $3.12 \times 10^{-4}$ |
| 650 | $2.78 \times 10^{-4}$ | $2.71 \times 10^{-4}$ |
| 700 | $2.70 \times 10^{-4}$ | $2.71 \times 10^{-4}$ |

Table 3: Training epochs and their respective RMSE

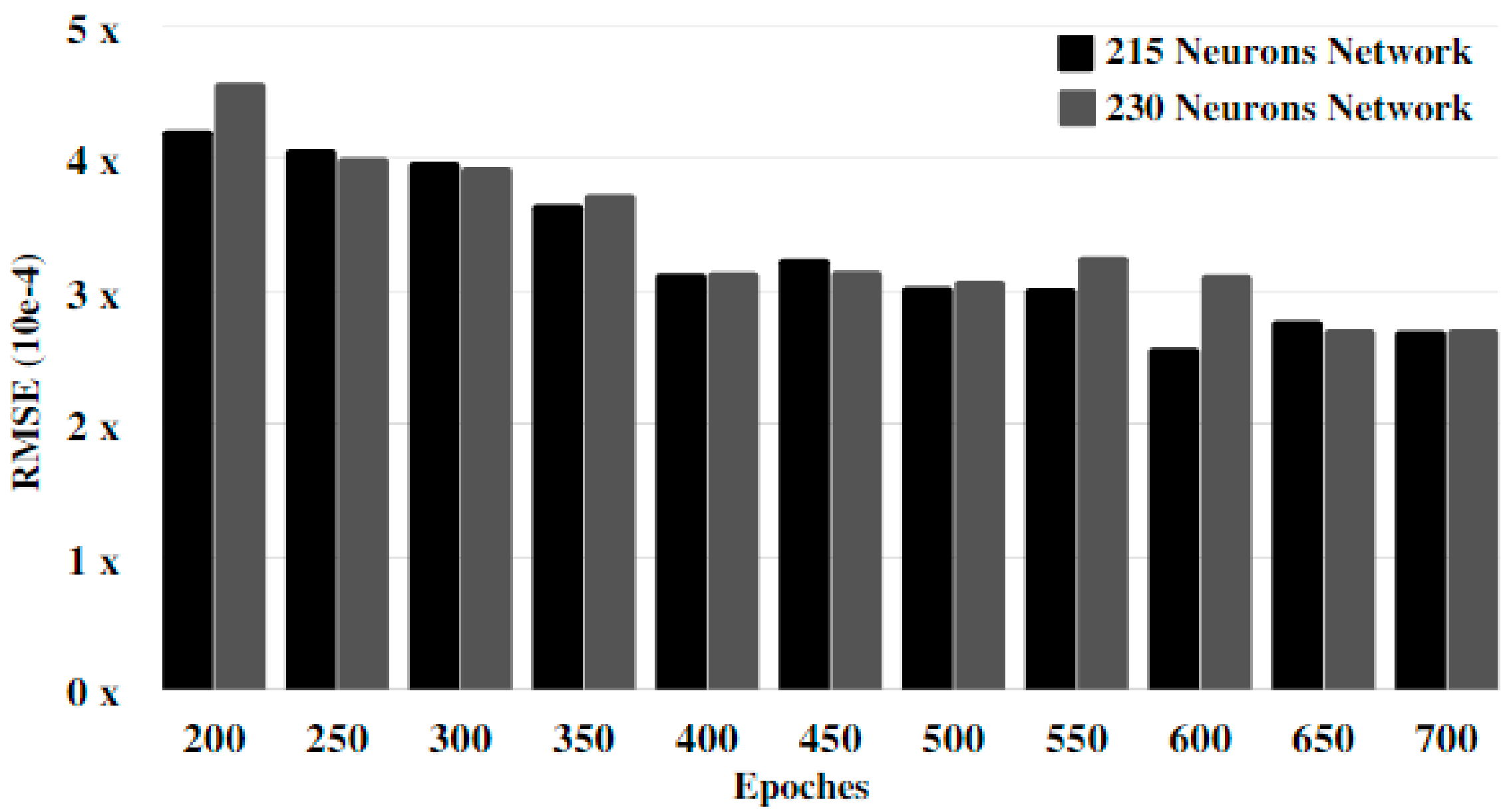

Figure 4: Training epochs and their respective RMSE

| Epoches | Training Time (hh:mm:sec) | |
|---|---|---|
| | **215** | **230** |
| 200 | 0:02:53.208191 | 0:02:50.031049 |
| 250 | 0:03:26.684626 | 0:03:29.684626 |
| 300 | 0:04:20.262378 | 0:04:06.984863 |
| 350 | 0:05:02.272071 | 0:04:56.698850 |
| 400 | 0:05:48.006247 | 0:05:35.373864 |
| 450 | 0:06:43.154403 | 0:07:20.691392 |
| 500 | 0:07:11.066267 | 0:07:44.962007 |
| 550 | 0:08:14.473178 | 0:07:49.877603 |
| **600** | **0:08:58.040198** | 0:08:36.912836 |
| 650 | 0:09:50.271088 | 0:09:34.858372 |
| 700 | 0:11:31.711259 | 0:11:13.938162 |

Table 4: Training epochs and their respective training time.

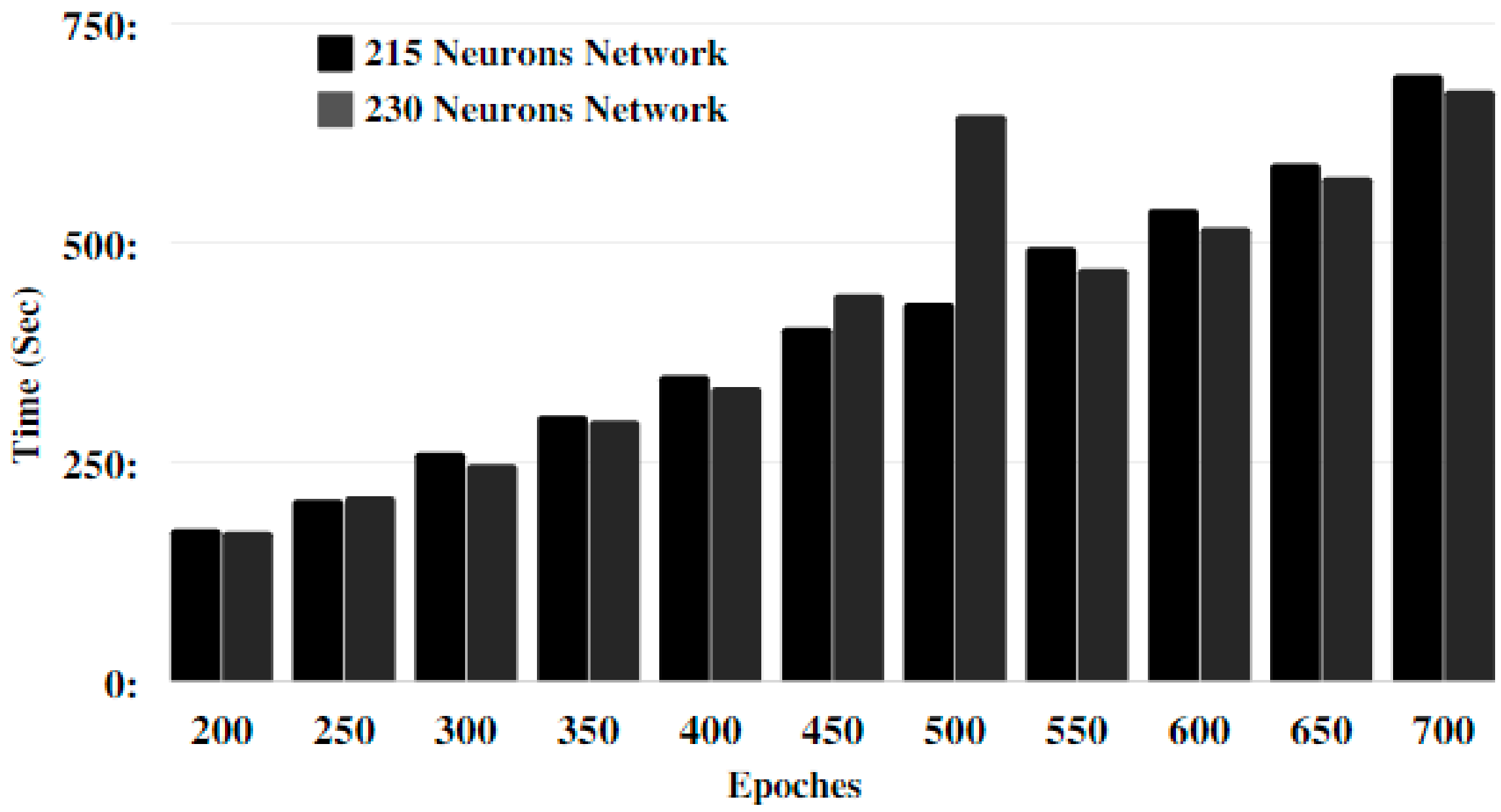

Figure 5: Training epochs and their respective training time.

The best combination of total number of neurons in the hidden layer and the total number epochs for the model to train with was in the case of 215 neurons and 600 epochs. In this case the RMSE had the least value of $2.57 \times 10^{-4}$. The time taken by this model for training was 0:08:58.040198 (hh:mm:sec).

## 4.1.3. Selection Of Model With Least RMSE

The table below shows the RMSE with different values of ‘k’ for the KNN model.

| ‘K’ Value | RMSE |
|---|---|
| 0 | 0.090962216036 12045 |
| 1 | 0.0011073469169303343 |
| 2 | 0.0010082267387196864 |
| 3 | 0.000979276752129196 |
| 4 | 0.0009693373511112318 |
| 5 | 0.0009670480106215172 |
| 6 | 0.0009673800261094776 |
| 7 | 0.0009682260068573489 |
| 8 | 0.0009697606728110148 |
| 9 | 0.0009722453685170264 |

Table 5: Different ‘k’ values and their respective RMSE

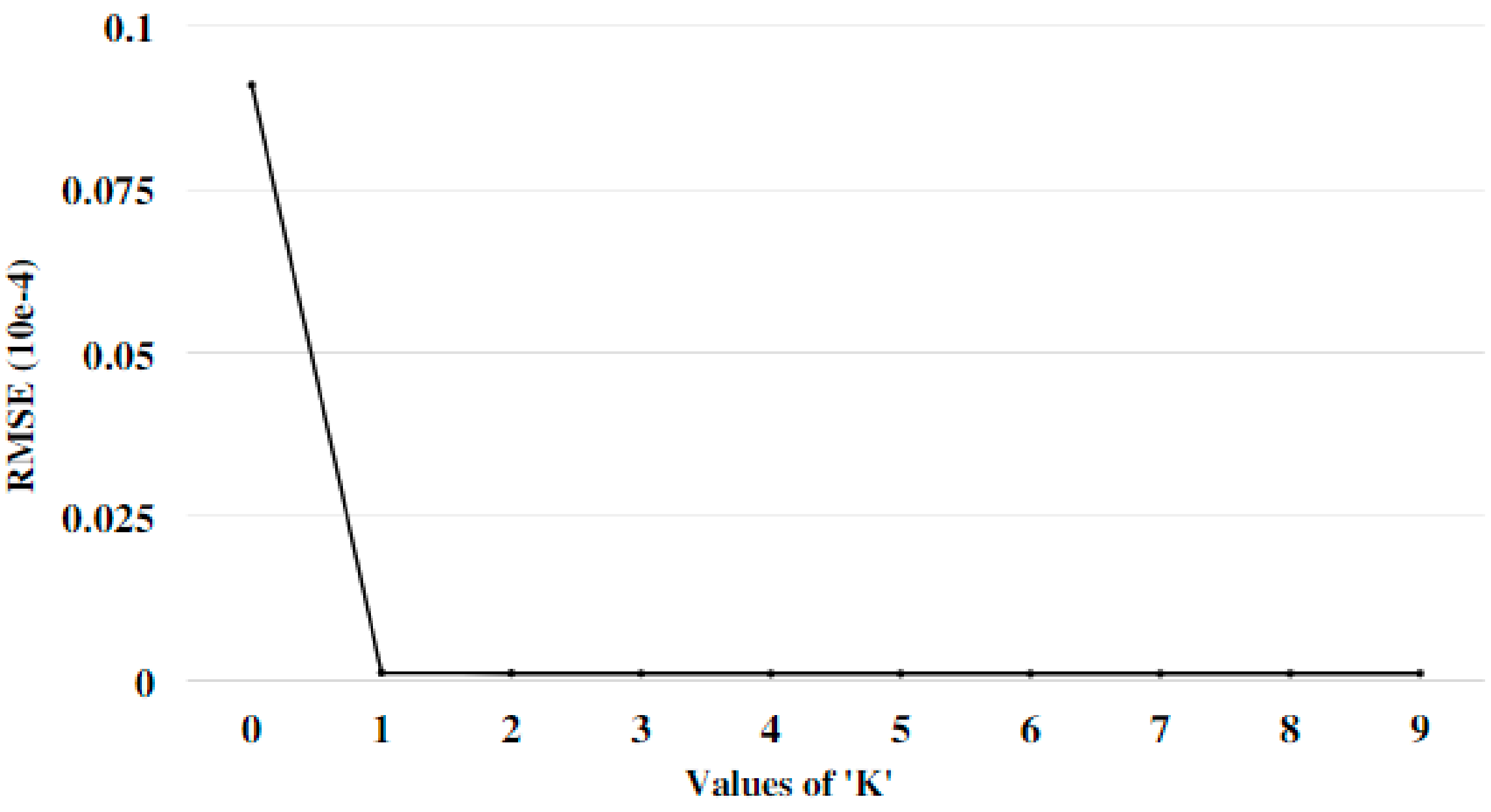

Figure 6: Different ‘k’ values and their respective RMSE

The least value of RMSE was obtained in the case of ‘k’ equal to 5. The value of RMSE was 9.67 x $10^{-4}$. However, this value of RMSE is way larger than the one obtained using the CNN model. Hence, the CNN model is preferred over KNN for this project.

Below is the plot for the predicticted values and their respective ground truths for the test instance of 1000 and bus number of 1 to 50.

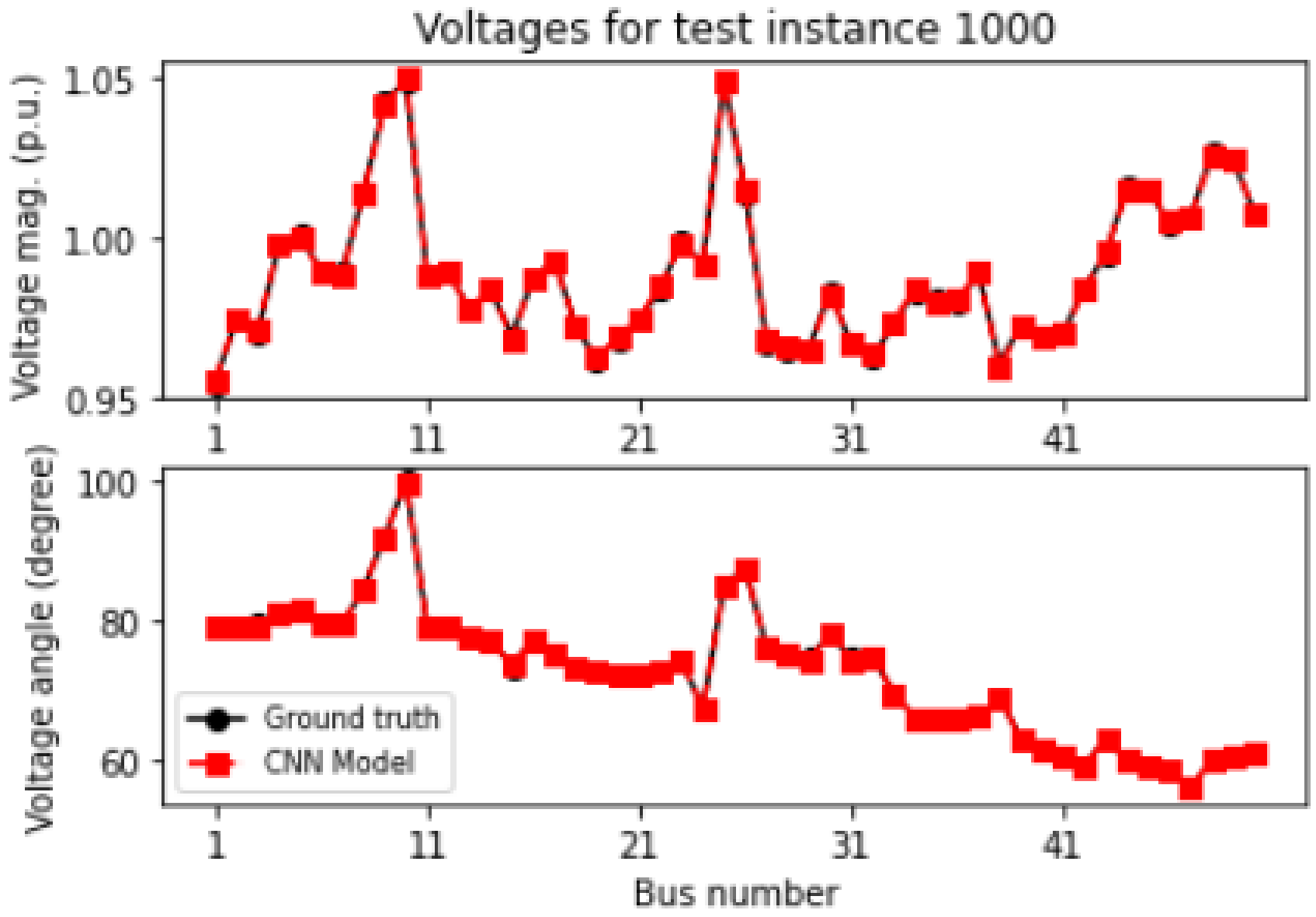

Figure 7: Estimation results and respective ground truth for 50 buses.

The paper [32] proposed a Prox-Linear Neural Network which predicted the RMSE of $2.97 \times 10^{-4}$ on the same data. The proposed neural network model in the paper is more complex and has a significantly higher number of hidden layers than the CNN neural network model proposed in this project work. Moreover, The CNN model also predicted results with lesser RMSE value than paper mentioned above. The CNN model with 215 neurons in a hidden layer with 600 epochs of training predicted RMSE of 2.57 x $10^{-4}$.

# 4.2. State Forecasting Using Deep Recurrent Neural Network

## 4.2.1. LSTM Model

Long Short-Term Networks (LSTM) unit is also a recurrent unit that has cyclic connections. The only difference between an LSTM unit and an RNN and unit is that a LSTM unit is more sophisticated and it is composed of gates. These gates I used to regulate the flow of information through the unit in a much better way in comparison to RNN.

A simple LSTM model was created with just one LSTM layer and an output layer. The model produced an RMSE of 1.18 x $10^{-2}$.The following were the graphs that were produced using the LSTM model.

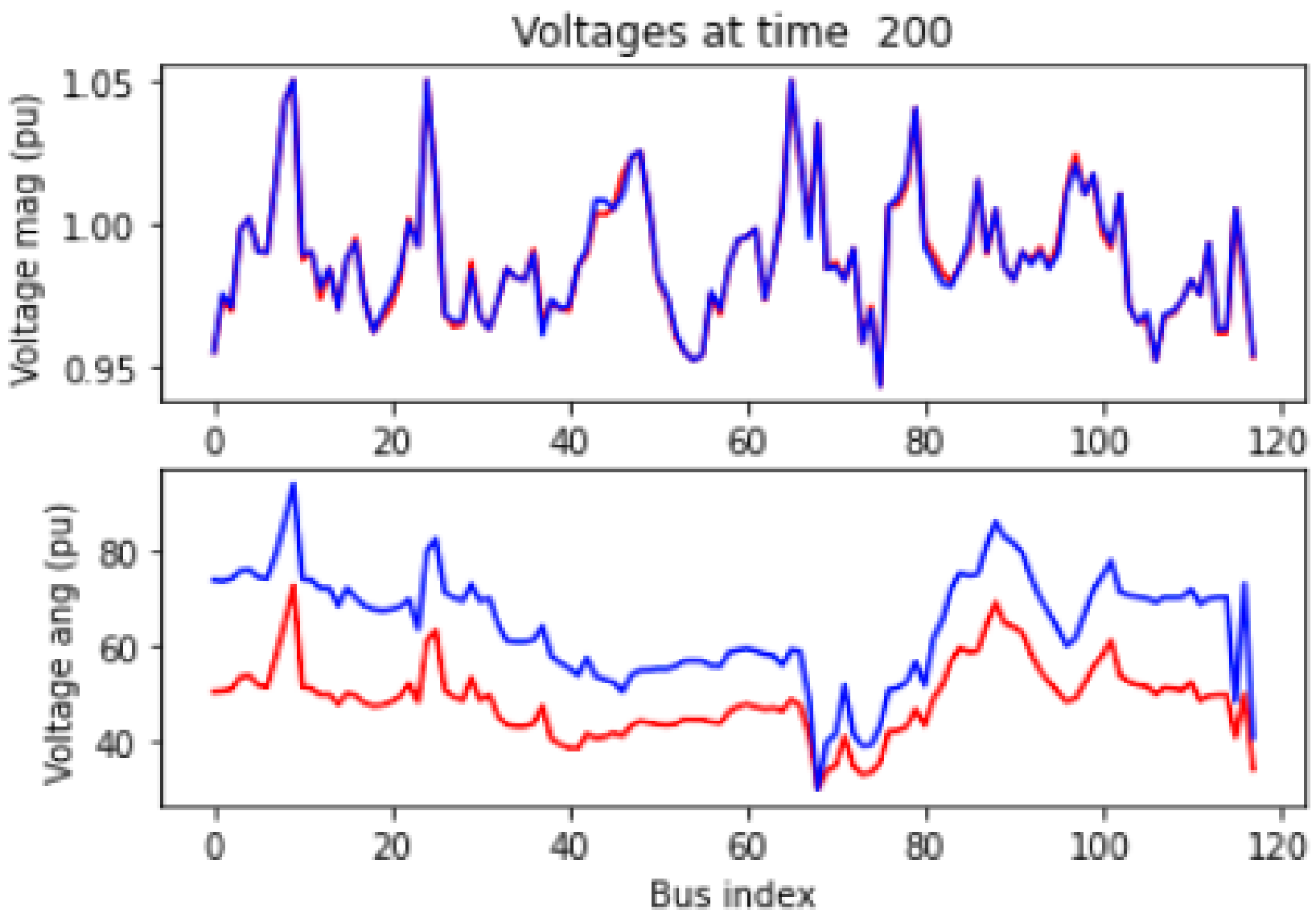

Figure 8: Plots for LSTM Model

The red line is for the ground truth values and the blue is for forecasted values. It could be noticed that though the forecasted voltage values are closer to the true values, the phase angle values have achieved a significant variation from the true values.

# 4.2.2. Single Time Distributed Layer Model

In this model a single 'Dense' layer iis given the feature of 'TimeDistributes' so that it can account for time sequences. The model only has a single time distributed dense layer, an LSTM layer with 20% dropout, and a final Dense layer for output.

The RMSE of this model is 1.17 x $10^{-2}$. The following were the graphs that were obtained from the model execution.

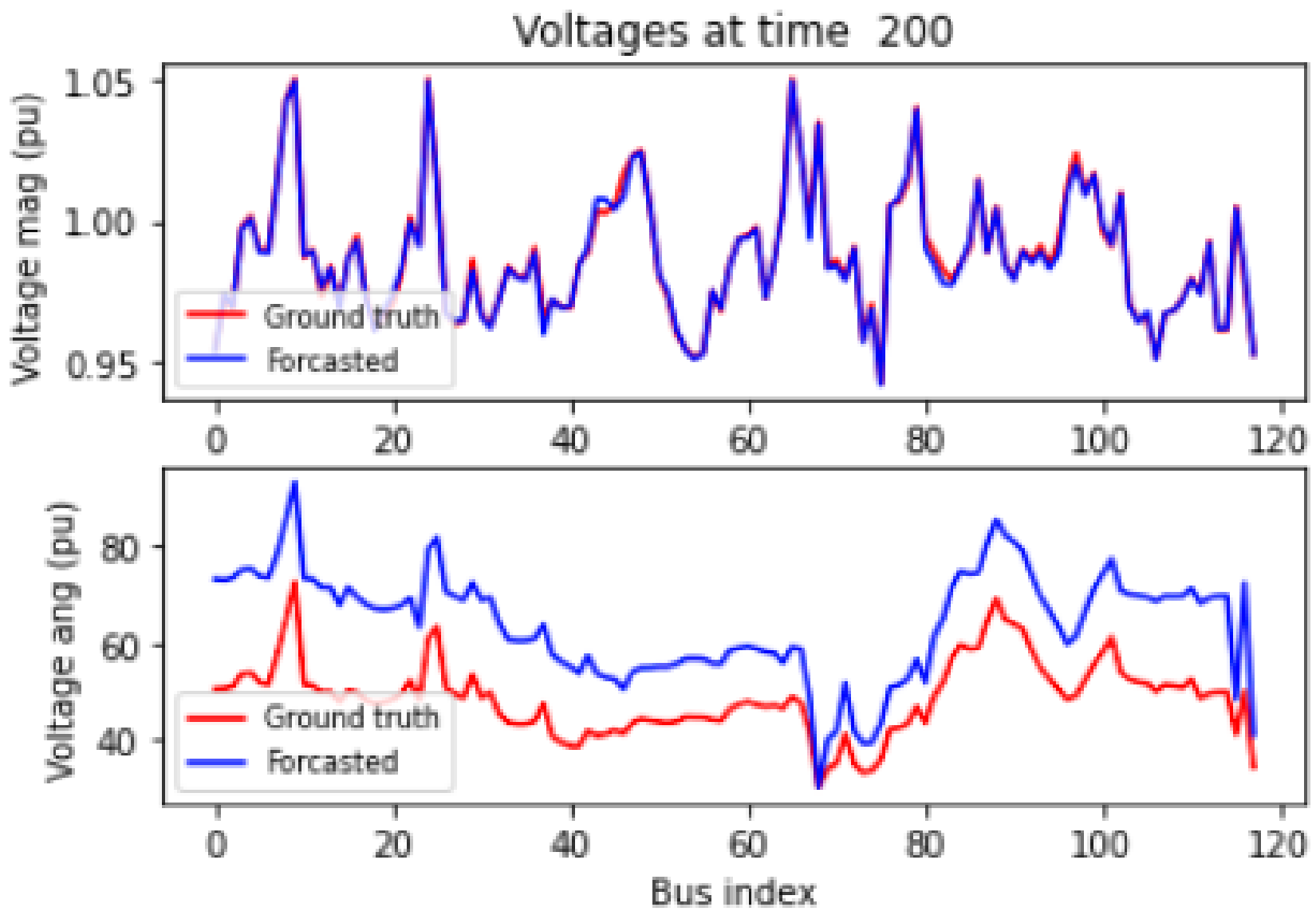

Figure 9: Plots for Single Time distributed layer Model

It can again be noticed that though the forecasted voltage values are closer to the true values, the phase angle values have achieved a significant variation from the true values.

## 4.2.3. Single SimpleRNN Layer Model.

This model has a single layer of 'SimpleRNN' and a dense layer for getting the outputs. The model does not use any LSTM layers. The RMSE obtained using the model was 1.17 x $10^{-2}$.

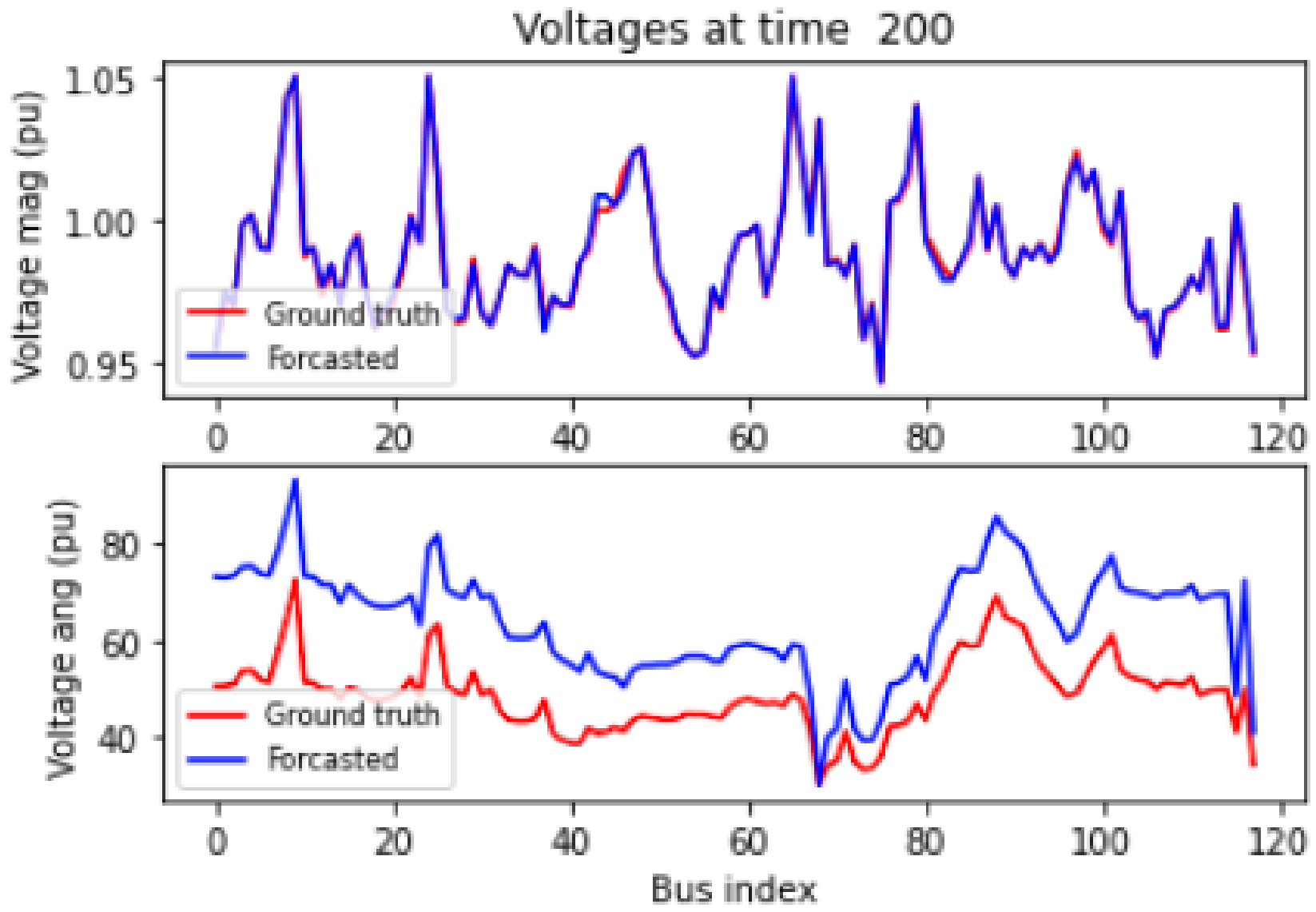

Figure 10: Plots for Single SimpleRNN layer Model

It can again be noticed that though the forecasted voltage values are closer to the true values, the phase angle values have achieved a significant variation from the true values.

### 4.2.4. Other Models

All the rest of models make use of SimpleRNN and/or Dense layers. These models do not need to be given a separate name and they have the same constituent elements. All the model structures can be obtained in the RNN_model.py file.

Model named 'stack_rnn_fase'

RMSE: $2.52 \times 10^{-3}$.

Graphs

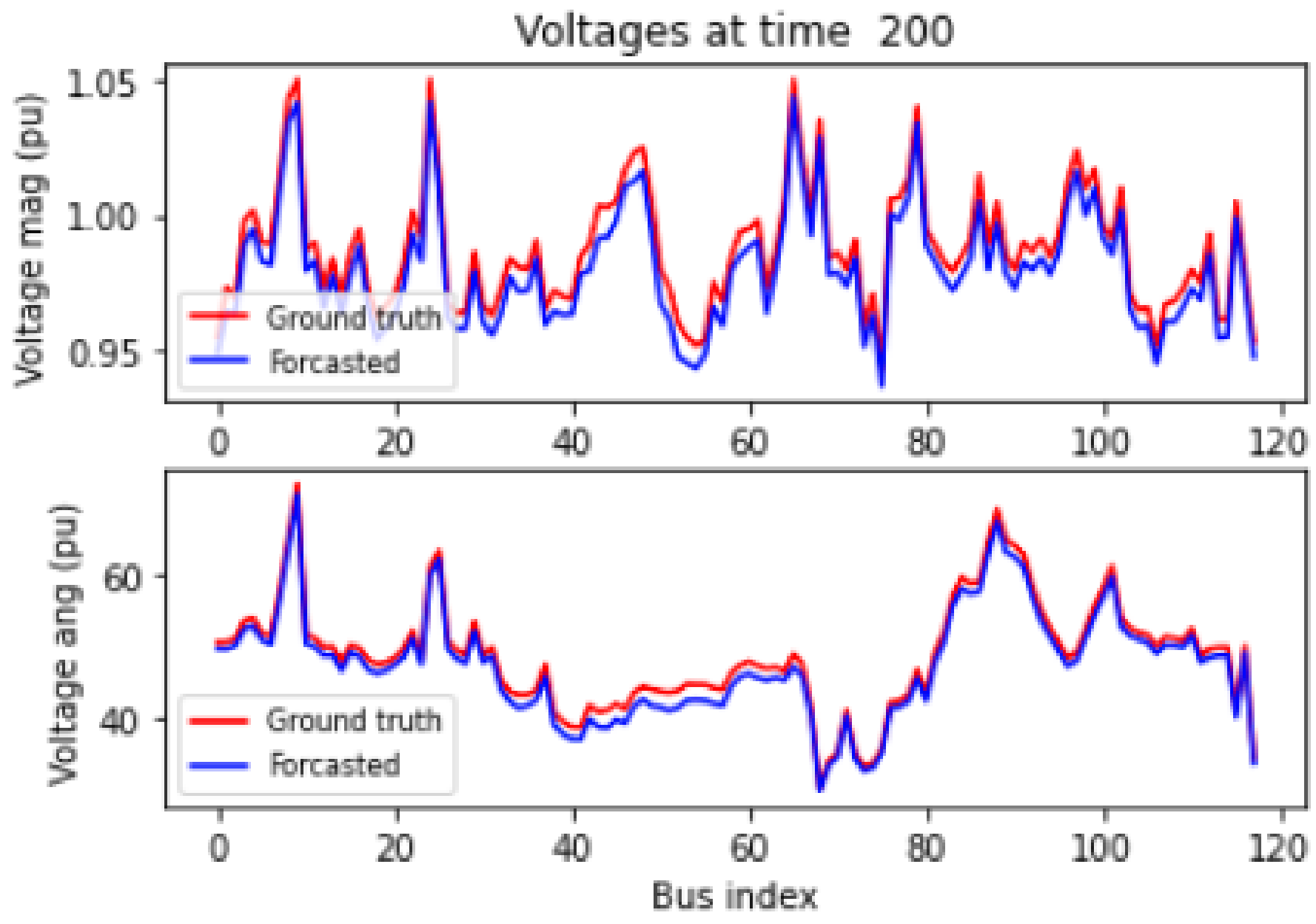

Figure 11: Plots for stack_rnn_fase Model

Model named ‘pretrained_rnn_plnet_fase’

RMSE: 2.9 x $10^{-3}$.

Graphs

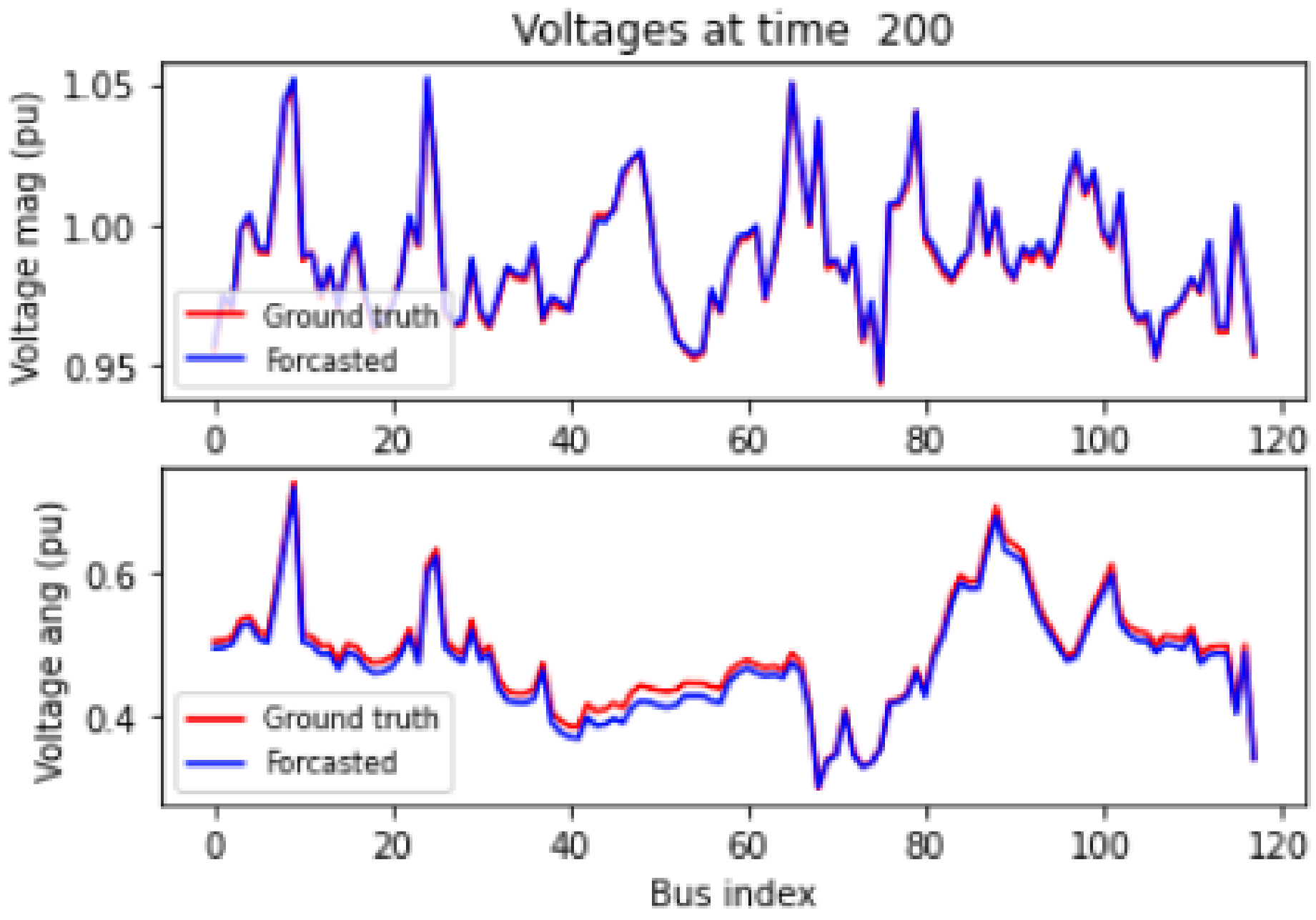

Figure 12: Plots for pretrained_rnn_plnet_fase Model

Model named ‘rnn_plnet_fase’

RMSE: 3.49 x $10^{-3}$.

Graphs

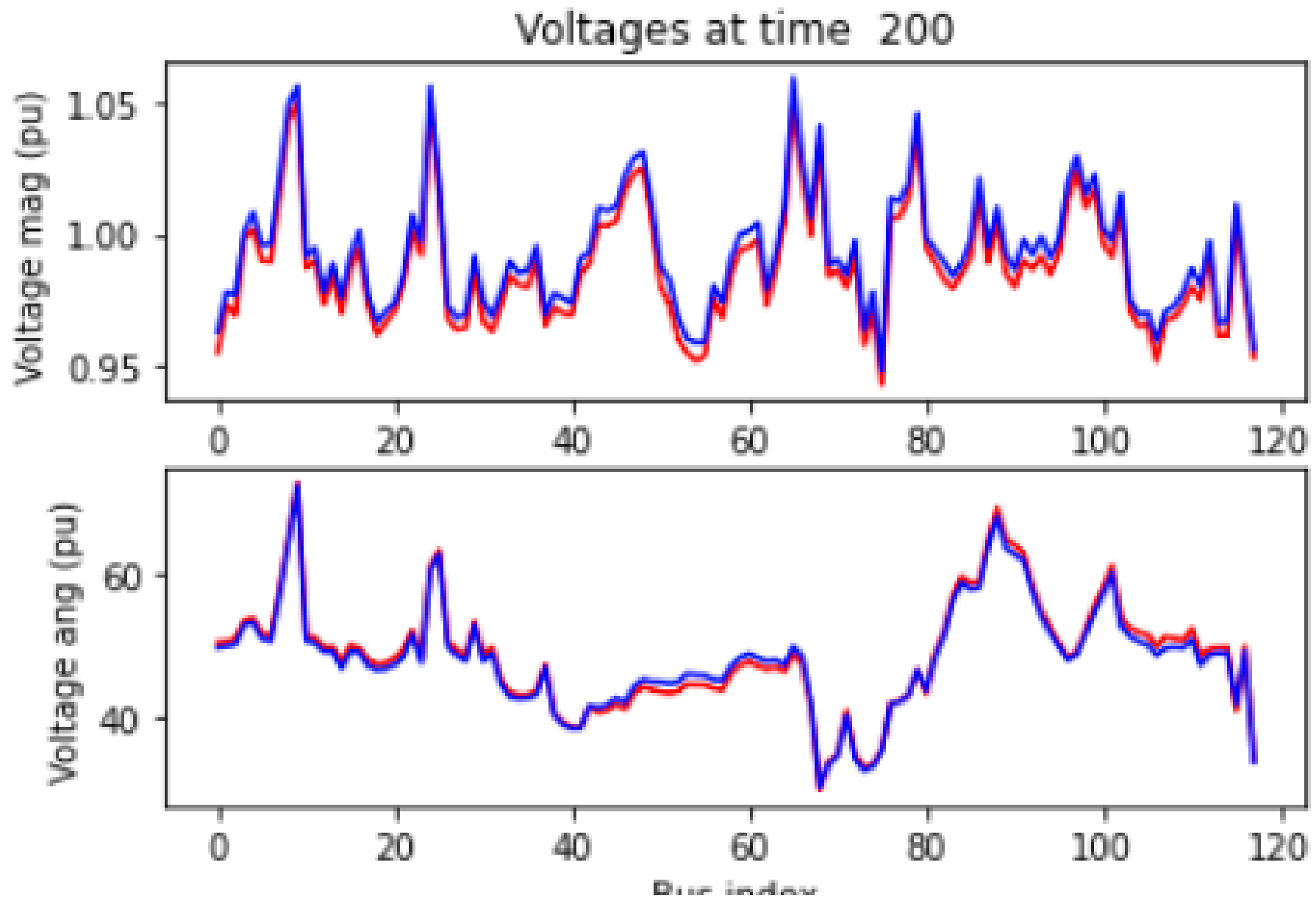

Figure 13: Plots for rnn_plnet_fase Model

Model named ‘simplified_rpln_fase’

RMSE: 1.18 x $10^{-2}$

Graphs

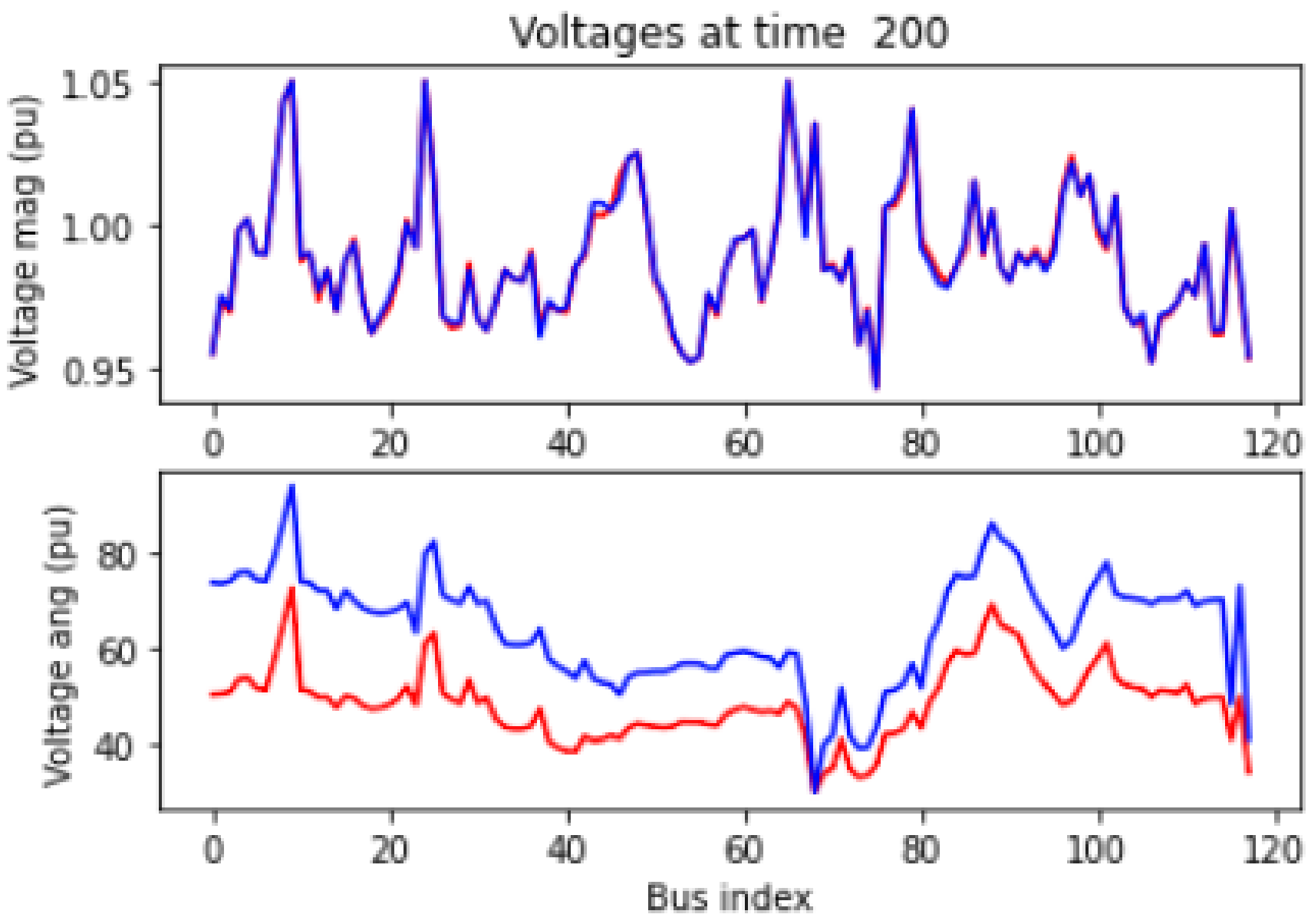

Figure 114: Plots for simplified_rpln_fase Model

| Model Name | RMSE |
|---|---|
| LSTM Model | 1.18 x $10^{-2}$ |
| Single Time distributed layer | 1.17 x $10^{-2}$ |
| Single SimpleRNN layer model | 1.17 x $10^{-2}$ |
| stack_rnn_fase | 2.52 x $10^{-3}$ |
| pretrained_rnn_plnet_fase | 2.9 x $10^{-3}$ |
| rnn_plnet_fase | 3.49 x $10^{-3}$ |
| simplified_rpln_fase | 1.18 x $10^{-2}$ |

Table 6: Various RNN models with respective RMSE.

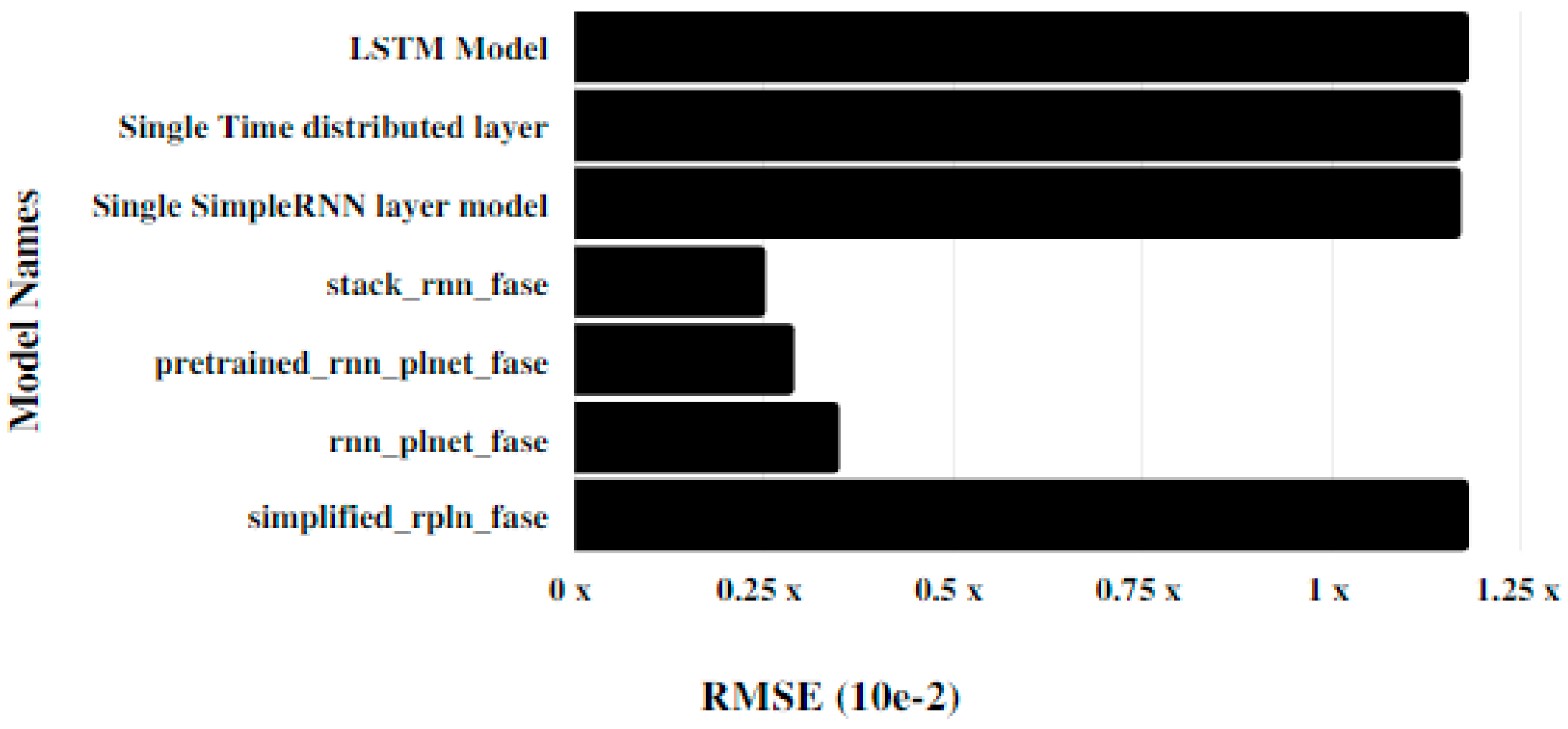

Figure 15: Various RNN models with respective RMSE.

# Chapter 5: Conclusion

To conclude, it was found that the existing Power System State Estimation (PSSE) techniques are little expensive in terms of computational costs and hence it necessitates the need for better state estimation techniques. For this problem, this research proposed a real-time state estimation technique for power grids using Convolutional Neural Networks. The research made use of SCADA measurements from 188 IEEE bus systems. The research made different CNN models with different numbers of neurons in layers and trained them for different numbers of epochs. It was found that the model with 215 neurons and trained for 600 epochs was able to produce the least amount of RMSE of $2.57 \times 10^{-4}$, which was comparatively accurate than one of previous studies involved in making the estimation using a Prox Linear Model ($2.97 \times 10^{-4}$).

Furthermore, the research also proposed a method for Power System State Forecasting for improving system awareness and resilience. The forecasting was carried out using a model of Recurrent Neural Network (RNN). This model helps in accounting for long-term nonlinear aspects present in data and based on that it does the forecasting. The model took the same data for forecasting. The model also made different models of RNN network and then proposed the model with the least RMSE value. The proposed model forecasted with a RMSE of 2.53 x $10^{-3}$, which is comparatively equal to the previous study mentioned above (2.59 x $10^{-3}$).

## 5.1. Future Work

As a part of the future work that one could do with this project, is to conduct the same or similar experiment with PMU (Phase Measurement Units) and SCADA data altogether. This could help in advancing the project with better and accurate results of prediction. Moreover, one could also look into the outliers and anomalies for the data and design a filtering model to clear out any such irregularities. The state estimator and the forecasting network could be made to work with live and real time data as well.

Moreover, the system could also be made more resilient to any cyber attacks by developing deep neural networks that detect such threats and try to prevent it. There are also works in considering other factors for estimation like environmental consideration. So such variables could be listed down and the data for them can be collected and the estimation could be made for better efficiency of the results.